\documentclass[]{aa}
\usepackage[varg]{txfonts}
\usepackage{graphicx}
\usepackage{multirow}
\usepackage{color}
\usepackage{hyperref}
\usepackage{subfigure}
\usepackage{wrapfig}
\begin{document}

\title{Spectral Gradient of the Thermal Millimetre Continuum as a Diagnostic for Optical Thickness in the Solar Atmosphere}

\author{A.S. Rodger \and N. Labrosse}

\institute{SUPA School of Physics and Astronomy, University of Glasgow, Glasgow, Scotland, G12 8QQ}

\abstract{}
{In this article we aim to show how the gradient of the thermal millimetre continuum spectrum, as emitted from the quiet solar atmosphere, may be used as a diagnostic for the optical thickness regime at the centre of the observing frequency band.}
{We show the theoretical derivation of the gradient of the millimetre continuum for both logarithmic- and linear-scale spectra.
We compare this expression with the empirical relationship between the slope of the millimetre continuum spectrum and the plasma optical thickness computed from both isothermal and multi-thermal two-dimensional cylindrical radiative transfer models.}
{It is found that the logarithmic-scale spectral gradient provides a clear diagnostic for the optical thickness regime for both isothermal and multi-thermal plasmas, provided that a suitable correction is made for a non-constant gaunt factor over the frequency band. 
For the use of observers we present values for this correction at all ALMA bands and at a wide range of electron temperatures.}
{We find that the spectral gradient can be used to find (a) whether the source is fully optically thin, (b) the optical thickness of the source if it lies within the transitional regime between optically thin and thick plasma ($\tau \approx 10^{-1} - 10^{1}$), or (c) whether the source is fully optically thick for an isothermal plasma.
A multi-thermal plasma will act the same as an isothermal plasma for case (a), however, the transitional regime will only extend from $\tau \approx 10^{-1} - 10^{0}$.
Above $\tau=1$ the slope of the continuum will depend increasingly on the temperature gradient, as well as the optical thickness, reducing the reliability of the diagnostic.}


\titlerunning{Millimetre Continuum Gradient as a Diagnostic for Optical Thickness}

\authorrunning{Rodger \& Labrosse}

\maketitle

\section{Introduction}\label{sec:intro}

Millimetre continuum emission has frequently been presented as a powerful, direct temperature diagnostic for the  plasma of the quiet solar atmosphere. 
Thanks to a predominantly collisional emission mechanism and lying within the Rayleigh-Jeans limit, there results a direct relationship between plasma electron temperature and millimetre-wavelength brightness temperature for a particular formation region within a line-of-sight (LOS) through an optically thick medium.
Observations using the millimetre regime have long been handicapped by low resolutions.
However, with the advent of the \emph{Atacama Large Millimeter/sub-millimeter Array} (ALMA) solar physicists have the ability to observe the solar atmosphere in the millimetre regime at unprecedented spatial resolutions. 
Solar science with ALMA has been discussed in several articles e.g. \cite{Heinzel2015,Wedemeyer2016, White2017, Shimojo2017, Rodger2017}.

The limitations of the millimetre brightness temperature as a plasma diagnostic occur when the optical thickness of the source in consideration is unknown.
In this article we aim to show how the gradient, or slope, of the millimetre continuum spectrum, across the observing band, may be used to estimate the optical thickness regime at the band centre. 
With the knowledge of the optical thickness regime at band centre it becomes possible to judge which plasma diagnostic is available from the brightness temperature: if it is clear the plasma is optically thick, the brightness temperature may be used as the direct diagnostic for the local electron temperature of the forming region. 

We show the derivation of the relationship between the gradient of the spectrum and the optical thickness of the emitting material for both a linear- and a logarithmic-scale spectrum discussing the merits of both as plasma diagnostics. 
The relationship between the gradient and optical thickness is tested using results from the 2-dimensional, cylindrical cross-section non-local thermodynamic equilibrium numerical radiative transfer models of \cite{Gouttebroze_Labrosse}. 

In Section~\ref{sec:theory} we present the theoretical derivation of the spectral gradient expressions used in this article. 
Section~\ref{sec:modelling} compares the results from both isothermal and multi-thermal numerical models to the derived expressions. 
The conclusions from this study are presented in Section \ref{sec:conclusions}.

\section{Theory}\label{sec:theory}

The millimetre spectrum in the quiet solar atmosphere is dominated by free-free collisional processes. 
The dominant emission mechanism across the millimetre regime is thermal bremsstrahlung, and thus we assume a purely thermal bremsstrahlung emission mechanism for the optical thickness diagnostic discussed in this letter.
The statement that thermal bremsstrahlung is dominant, however, becomes untrue at low temperatures, below 5000~K, and at high densities where neutral hydrogen absorption becomes significant \citep{Rutten2017}, and thus due care should still be given when using these diagnostics.
The frequency-dependent absorption coefficient, $\kappa_{\nu}$, for thermal bremsstrahlung is described by \citep{Dulk1985, Wedemeyer2016}:
\begin{equation}\label{eq:absorption_coefficient}
	\kappa_{\nu} = 9.78\times10^{-3}\frac{n_e}{\nu^{2}T^{3/2}}g_{\mathrm{ff}}\sum_i Z_i^2n_i, 
\end{equation}
in cgs units, where $n_e$ is the electron density, $T$ is the electron temperature, and $g_{\mathrm{ff}}$ is the thermal gaunt factor. 
$Z_i$ and $n_i$ are the charge and density for the ion species $i$. 
The optical thickness of an homogeneous line-of-sight (LOS) of length $L$ can be approximated as $\tau_{\nu} = \kappa_{\nu}L$, such that the optical thickness, when assuming purely thermal bremsstrahlung absorption, will vary with frequency as $\tau_{\nu} \propto g_{\mathrm{ff}}\nu^{-2}$, where $g_{\mathrm{ff}}$ is slowly varying with frequency and temperature. 

The advantage of observing the solar atmosphere in the millimetre regime is the strong potential for temperature diagnostics.
This statement arises from the emission mechanism being dominated by thermal mechanisms such that the source function of the emission can be described by the Planck function, whilst the millimetre regime also lies within the Rayleigh-Jeans limit.
Combining these two aspects of solar millimetre emission, the brightness temperature can be described as:
\begin{equation}\label{eq:brightness_temperature}
	T_{\mathrm{B}}(\nu) = \int T \kappa_{\nu} \mathrm{e}^{-\tau_{\nu}} \mathrm{d}s \ ,
\end{equation}
where the integration is performed over a LOS of length $s$ with a path element of $\mathrm{d}s$.
For a sufficiently optically thick source the brightness temperature will tend towards saturating at the electron temperature of the plasma. 
For this diagnostic to be used successfully, however, knowledge of the source's optical thickness at the observing wavelength is required, as optically thin material, or not-sufficiently optically thick material will provide brightness temperatures non-representative of the electron temperature. 

In this section we discuss how the gradient of the logarithmic brightness temperature spectrum may be used as a diagnostic for the optical thickness at the centre of the observing band. 
By differentiating the logarithm of Eq.~(\ref{eq:brightness_temperature}) with respect to the logarithm of the frequency the gradient of the logarithmic spectrum is found to be:
\begin{equation}\label{eq:loglog_general}
	\frac{\mathrm{d\,log}(T_{\mathrm{B}})}{\mathrm{d\,log}(\nu)} = \frac{\nu}{T_{\mathrm{B}}} \int T \mathrm{e}^{-\tau_{\nu}}(\frac{2}{\nu} - \frac{g'_{\mathrm{ff}}}{g_{\mathrm{ff}}})(\tau_{\nu}-1)\mathrm{d}\tau_{\nu}, 
\end{equation}
where $g'_{\mathrm{ff}}$ is the rate of change of the thermal gaunt factor with frequency. 

If it can be assumed that the LOS is isothermal and $g'_{\mathrm{ff}} \approx 0$, Eq.~(\ref{eq:loglog_general}) can be simplified to:
\begin{equation}\label{eq:loglog_isothermal}
	\frac{\mathrm{d\,log}(T_{\mathrm{B}})}{\mathrm{d\,log}(\nu)} = \frac{-2\tau_{\nu}}{\mathrm{e}^{\tau_{\nu}}-1},
\end{equation}
such that the gradient of the logarithmic spectrum is dependent on the optical thickness of the source material solely.
In the high optical thickness limit, $\tau_{\nu} \gg 1$, Eq.~(\ref{eq:loglog_isothermal}) reduces to 0, whilst in the low optical thickness limit, $\tau_{\nu} \ll 1$ it reduces to -2. 
Thus by measuring the gradient of a small enough frequency band the optical thickness regime at band centre may be estimated. 


In the earlier part of this section we described how the gradient of the logarithmic  brightness temperature spectrum may be used as a diagnostic for the optical thickness at the centre of the observing band. 
For completeness, here we shall discuss the gradient of the linear expression for $T_{\mathrm{B}}(\nu)$. 
Similar to Eq.~(\ref{eq:loglog_general}) the general form for the brightness temperature spectral gradient is given by:
\begin{equation}\label{eq:standard_general}
	\frac{\mathrm{d}T_{\mathrm{B}}}{\mathrm{d}\nu} = \int T \mathrm{e}^{-\tau_{\nu}}(\frac{2}{\nu} - \frac{g'_{\mathrm{ff}}}{g_{\mathrm{ff}}})(\tau_{\nu}-1)\mathrm{d}\tau_{\nu}.
\end{equation} 
For an isothermal LOS and a thermal gaunt factor approximately constant with frequency, Eq.~(\ref{eq:standard_general}) simplifies to:
\begin{equation}\label{eq:standard_isothermal}
	\frac{\mathrm{d}T_{\mathrm{B}}}{\mathrm{d}\nu} = \frac{-2T\tau_{\nu}\mathrm{e}^{-\tau_{\nu}}}{\nu}.
\end{equation}
In the extreme optical thickness limits Eq.~(\ref{eq:standard_isothermal}) reduces to 0 for $\tau_{\nu} \gg 1$ and to $\frac{-2T(\tau_{\nu}-\tau_{\nu}^2)}{\nu}$ for $\tau_{\nu} \ll 1$ .
Hence for an optically thin source the linear-scale spectral gradient will vary with both frequency and temperature as well as optical thickness, such that the value of the gradient is non-unique for a given optical thickness. 
Hence due to the relative simplicity of the two diagnostics we conclude that the gradient of the logarithmic brightness temperature spectrum is a stronger optical thickness diagnostic than the gradient of the linear-scale brightness temperature spectrum.

\section{Modelling}\label{sec:modelling}

To test the theory presented in Section~\ref{sec:theory} we use a set of numerical radiative transfer models. 
The models we use are the two-dimensional, cylindrical cross-section, non-local thermodynamic equilibrium (non-LTE) radiative transfer prominence models of \cite{Gouttebroze_Labrosse}.
Each cylinder cross-section considers a mixture of hydrogen and helium at a range of isobaric pressures with either isothermal or multi-thermal radial temperature distributions.
The code considers a 34 level He atom, which includes 4 for He II and 1 for He III, as well as a 5 level plus continuum H atom.
By calculating the full non-LTE radiative transfer for each model we find the energy level populations and thus ionization densities for both atomic species. 
The cylinder is orientated such that its axis is parallel to the solar surface.  
We then calculate the brightness temperature for a set of horizontal lines-of-sight through the prominence cylinder. 
We follow the same methods for brightness temperature calculation as outlined in \cite{Rodger2017} with one notable change in the calculation of the thermal gaunt factor.
In this study we no longer estimate the thermal gaunt factor through the assumption given in \cite{Dulk1985} for plasma of temperature < $2\times 10^5$~K, but instead take our values by interpolating the table of calculated thermal gaunt factors of \cite{vanHoof}, as described in \cite{Gayet, Simoes}.
Whilst the optical thickness diagnostic discussed in this study assumes purely thermal bremsstrahlung absorption the results from these numerical models are calculated using both thermal bremsstrahlung and neutral hydrogen absorption.
For this analysis we first consider a set of isothermal prominence models to see how well they agree with Eq.~(\ref{eq:loglog_isothermal}) and see how this changes when considering a multi-thermal plasma instead.

Whilst nominally the models used here describe solar prominences, the results are applicable to any off-limb solar atmospheric structure. 
For on-disc structures, the spectral gradient will follow a similar relation to Eq.~(\ref{eq:loglog_isothermal}) with an additional term corresponding to the contribution of the background solar continuum spectrum.
On-disc structures would thus need knowledge of the brightness temperature of the structure and the background emission illuminating it from the solar disc. 
This will likely be problematic unless the structure in observation is formed above the chromosphere, the forming region for most millimetre radiation, or where the structure is transient in nature such that measurements of both the background and enhanced brightness temperature phases are obtainable.

\subsection{Isothermal Models}\label{sec:isothermal}

\begin{figure}
	\centering
	\includegraphics[width=\linewidth]{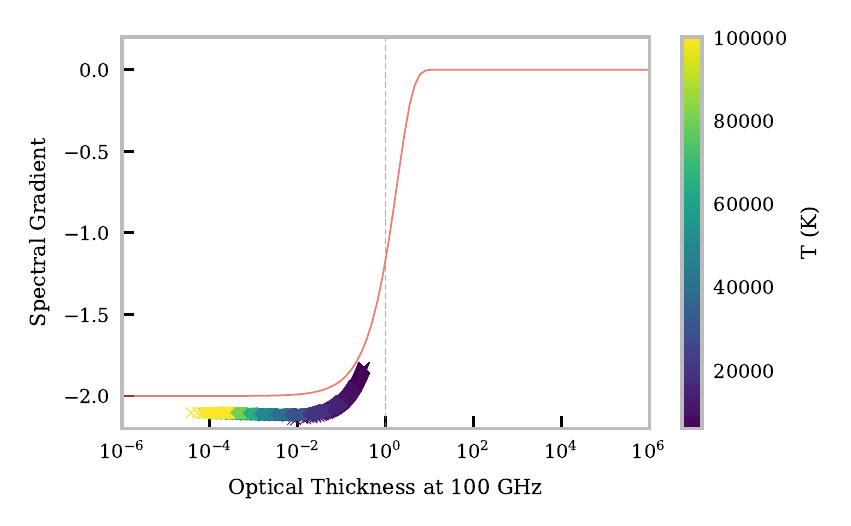}
	\caption{Relationship between optical thickness and logarithmic spectral gradient for a set of isothermal prominence models. 
		The solid red line shows the simple relationship as defined by Eq.~(\ref{eq:loglog_isothermal}).
	The dashed-blue line shows $\tau=1$.}
	\label{fig:isothermal_uncorrected}
\end{figure}

The set of isothermal models we use here are the same as the ``$t$'' models as described in \cite{Gouttebroze_Labrosse}.
They each consider a prominence cylindrical cross-section of radius 0.5 Mm at an altitude of 10 Mm. 
The isobaric pressure is 0.1 $\mathrm{dyn\,cm^{-2}}$, the helium abundance is 0.1 and the micro-turbulent velocity is 5~$\mathrm{km\,s^{-1}}$ in all models.
The set of  temperatures considered is \{6000, 8000, 10000, 15000, 20000, 30000, 40000, 50000, 65000, 80000, 100000\}~K.

For each isothermal model listed above we calculated the brightness temperature at the four spectral sub-bands of ALMA Band 3, i.e. 93, 95, 105 and 107 GHz \citep{White2017}.
From these brightness temperature values we fit a straight line to the logarithmic spectra of the millimetre continuum.
The plot showing the optical thickness of the LOS versus the logarithmic spectral gradient for all isothermal models is shown in Figure~\ref{fig:isothermal_uncorrected} alongside the simple derived expression from Eq.~(\ref{eq:loglog_isothermal}). 

From Figure~\ref{fig:isothermal_uncorrected} it can be seen that whilst the modelled spectral gradient -- optical thickness relationship follows a similar trend to that expected by Eq.~(\ref{eq:loglog_isothermal}) the values are slightly lower. 
This discrepancy is caused by the assumption that the gaunt factor is approximately constant over the frequency band.
In fact re-computing the brightness temperatures with a constant gaunt factor removed this discrepancy entirely. 

If the thermal gaunt factors variation across the observing band is to be significant enough to affect this relationship a method to account for it must be found. 
From Eq.~(\ref{eq:loglog_general}) it can be found for an isothermal plasma that; 
\begin{equation}\label{eq:correction}
	\frac{\mathrm{d\,log}(T_{\mathrm{B}})}{\mathrm{d\,log}(\nu)} = \frac{\mathrm{d\,log}(T_{\mathrm{B}})}{\mathrm{d\,log}(\nu)} \bigg\rvert_{g'= 0} \alpha,
\end{equation}
where $\alpha$ is the offset factor described by;
\begin{equation}\label{eq:alpha}
	\alpha = 1 - \frac{\nu g'_{\mathrm{ff}}}{2{g_{\mathrm{ff}}}}.
\end{equation}

\begin{figure}
	\centering
	\includegraphics[width=\linewidth]{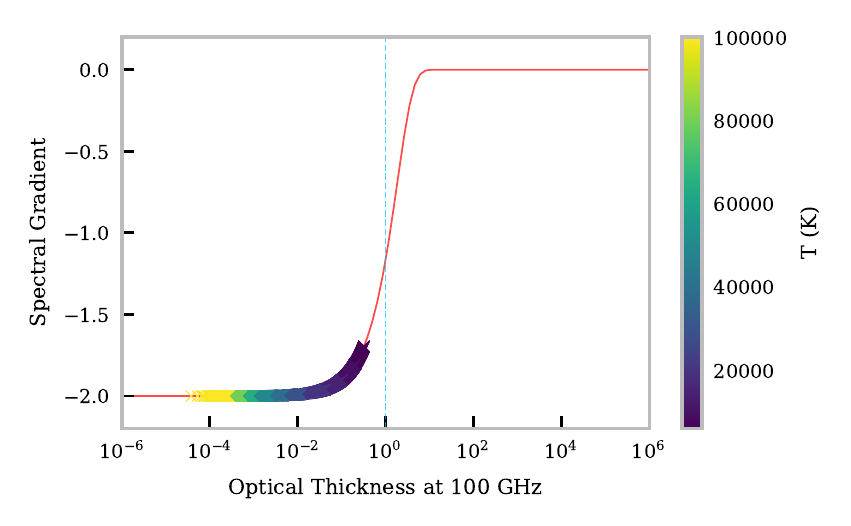}
	\caption{Same as Figure~\ref{fig:isothermal_uncorrected}, however the spectral gradient has been corrected using the known temperature of the isothermal models for the non-zero $g'_{\mathrm{ff}}$ through Eq.~(\ref{eq:correction}) \& Eq.(\ref{eq:alpha}).}
	\label{fig:isothermal_corrected}
\end{figure}

We evaluated $\alpha$ at the known isothermal temperatures of each prominence model and at the frequency of ALMA Band 3 centre (100~GHz). 
By dividing the modelled spectral gradient vs optical thickness relationship by this correcting factor we find results as shown in Figure~\ref{fig:isothermal_corrected}. 
It can be seen that as long as the non-zero rate of change of thermal gaunt factor with frequency is corrected for, isothermal models provide well the expected relationship between optical thickness and logarithmic spectral gradient as described in Eq.~(\ref{eq:loglog_isothermal}).

\subsection{Multi-Thermal Models}\label{multi-thermal} 

\begin{figure}
	\centering
	\includegraphics[width=\linewidth]{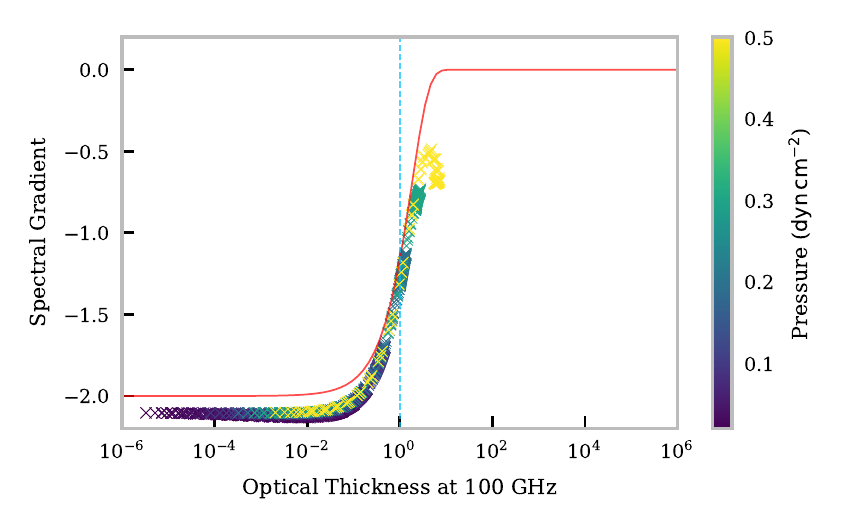}
	\caption{Same as Figure~\ref{fig:isothermal_uncorrected},  for a set of multi-thermal isobaric prominence models at various  pressures. }
	\label{fig:multithermal_uncorrected}
\end{figure}

In this section of this study we replicate the previous tests using multi-thermal prominence models. 
The prominence models which we use are the same as the '$p$' models as described in \cite{Gouttebroze_Labrosse}. 
The radius of these models is larger than that for the $t$ models at a value of 1 Mm. 
The non-isothermal temperature distribution is ad-hoc and defined by equation 1 of \cite{Gouttebroze2006}. 
Each model has a different isobaric pressure from the set of pressures; \{0.02, 0.03, 0.05, 0.10, 0.20, 0.30, 0.50\}~$\mathrm{dyn\,cm^{-2}}$.
The altitude, helium abundance and microturbulent velocity for all models are the same as described for the isothermal models. 

The same process as for the isothermal models was followed whereby the brightness temperature spectrum was calculated and the gradient was found for each of the sub-band wavelengths of ALMA Band 3. 
The relationship between logarithmic spectral gradient and optical thickness at the centre of Band 3 for this set of multi-thermal models can be seen in Figure~\ref{fig:multithermal_uncorrected}.

It can be seen again in Figure~\ref{fig:multithermal_uncorrected} that the relationship found in the simulated data is again below the simple relationship in Eq.~(\ref{eq:loglog_isothermal}) during the optically thin regime, as expected due to the same issues with non-zero rate of change of gaunt factor with frequency. 
However, when considering a set of multi-thermal LOSs, or generally a structure of unknown temperature, it is less simple to correct for $\alpha$ through Eq.~(\ref{eq:correction}) and Eq.~(\ref{eq:alpha}) due to the lack of a single representative temperature value. 
To attempt to find a solution to this, we have calculated $\alpha$ at all ALMA Bands and at a wide range of temperatures between $10^3$ and $10^6$~K.

\begin{figure}
	\centering
	\includegraphics[width=\linewidth]{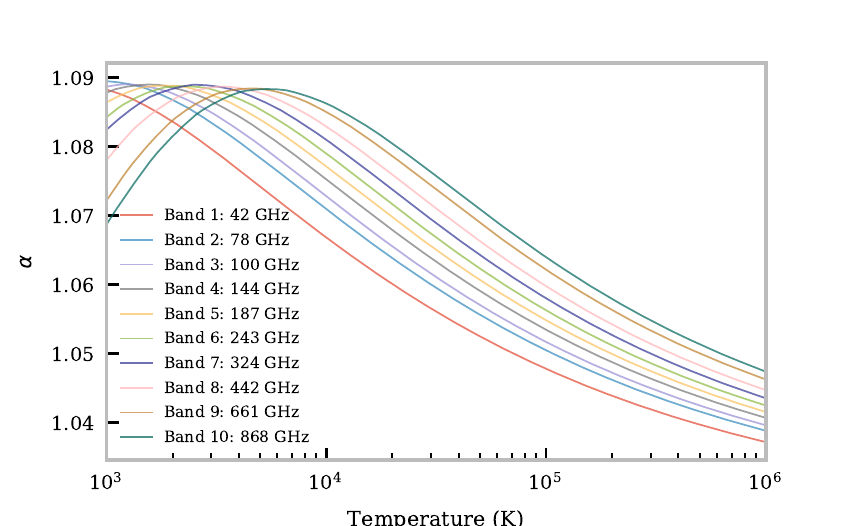}
	\caption{Smoothed variation of $\alpha$ correction, evaluated at all ALMA bands over a wide range of temperatures.}
	\label{fig:alpha}
\end{figure}

\begin{figure}
	\centering
	\includegraphics[width=\linewidth]{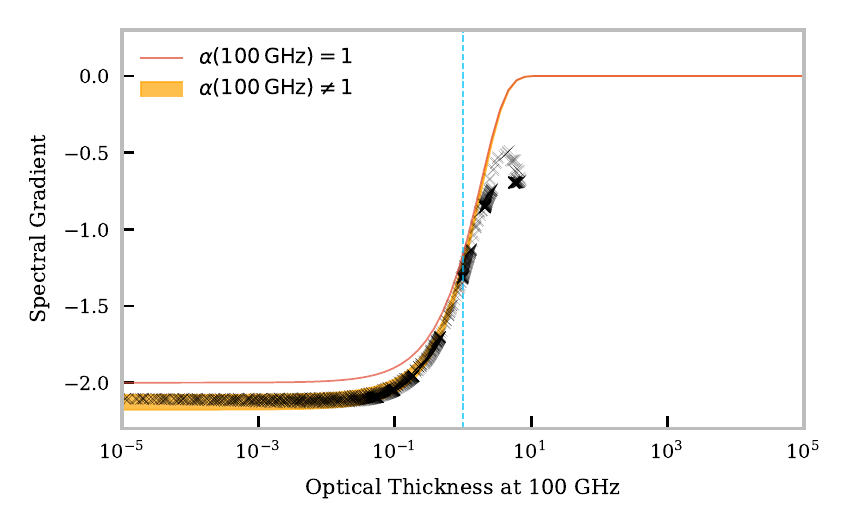}
	\caption{The relationship between optical thickness and spectral gradient for all multi-thermal numerical models is shown in black. 
		The simple, isothermal expression from Eq.~(\ref{eq:loglog_isothermal}), without the correction for $\alpha$ factor is shown in red. 
		The orange region shows the corrected relationship from Eq.~(\ref{eq:correction}), where $\alpha$ is evaluated for ALMA Band 3 at temperatures between $10^3$ and $10^6$~K.
	The dashed-blue line shows $\tau=1$.}
	\label{fig:multithermal_corrected} 
\end{figure}

Due to the very small magnitude of the rate of change of the gaunt factor with frequency ($g^{\prime}_{\mathrm{ff}} \sim 10^{-13}$), the $\alpha$ variation with temperature displayed a jagged, oscillation-like pattern at low temperatures.
This numerical artifact was removed by fitting $g^{\prime}_{\mathrm{ff}}$ with the function;
\begin{equation}\label{eq:g_dashed_fit}
	g^{\prime}_{\mathrm{ff}}(\nu,T) = a(\nu)\frac{T^{b(\nu)}}{T^{c(\nu)}+d(\nu)}, 
\end{equation}
and calculating $\alpha$ using the fitted values for $g^{\prime}_{\mathrm{ff}}$.
$a(\nu)$, $b(\nu)$, $c(\nu)$ and $d(\nu)$ are constants dependent on the frequency band.
The resulting smoothed variation of $\alpha$ for each ALMA Band with temperature is shown in figure~\ref{fig:alpha}.
The temperature values in figure~\ref{fig:alpha} extend below the temperature range ($\sim$ 5000~K) where this method may be applied, as neutral hydrogen will become a more significant emission mechanism.
%
%
Applying the maximum and minimum values for $\alpha(100\mathrm{GHz})$ to Eq.~(\ref{eq:correction}) we compare the corrected relationship to the results from the multi-thermal models.
The results from this method are shown in Figure~\ref{fig:multithermal_corrected}.

Figure~\ref{fig:multithermal_corrected} shows that the $\alpha$(100~GHz) correction can produce a fairly close agreement to the values found from the multi-thermal numerical models, although the relationship notably differs at higher optical thickness.
This is primarily due to the breakdown in assumption that the LOS is isothermal as the spectral gradient becomes dependent on the temperature gradient of the LOS in addition to the optical thickness. 
Using the non-zero $g'_{\mathrm{ff}}$ corrected logarithmic spectral gradient may, however, be used to discern whether the emission is (a) optically thin, (b) the optical thickness of the material if it is in the range $\tau \approx 10^{-1} - 1$, or (c) whether it is optically thick and the gradient is defined by the temperature gradient of the plasma. 

\subsection{Minimum Required Uncertainty in Brightness Temperature Measurement}

\begin{figure*}[t!]
	\centering
	\begin{subfigure}
		\centering
		\includegraphics[width=0.45\linewidth]{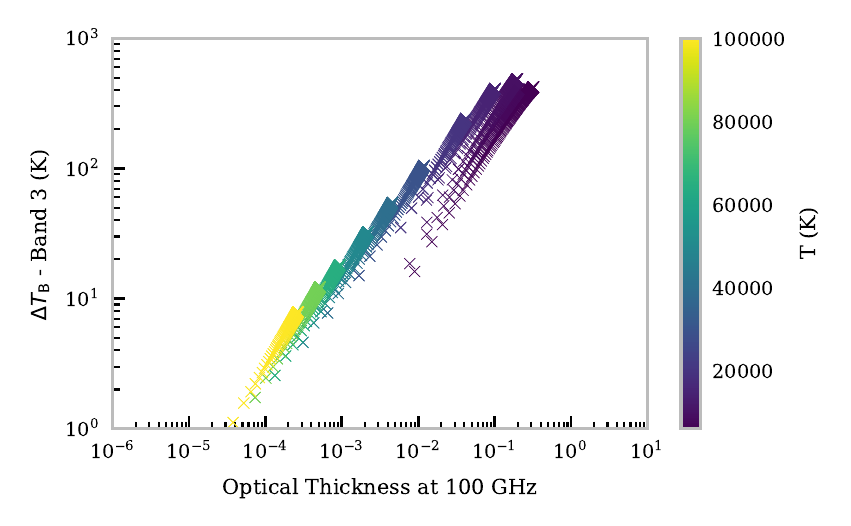}
	\end{subfigure}
	~
	\begin{subfigure}
		\centering
		\includegraphics[width=0.45\linewidth]{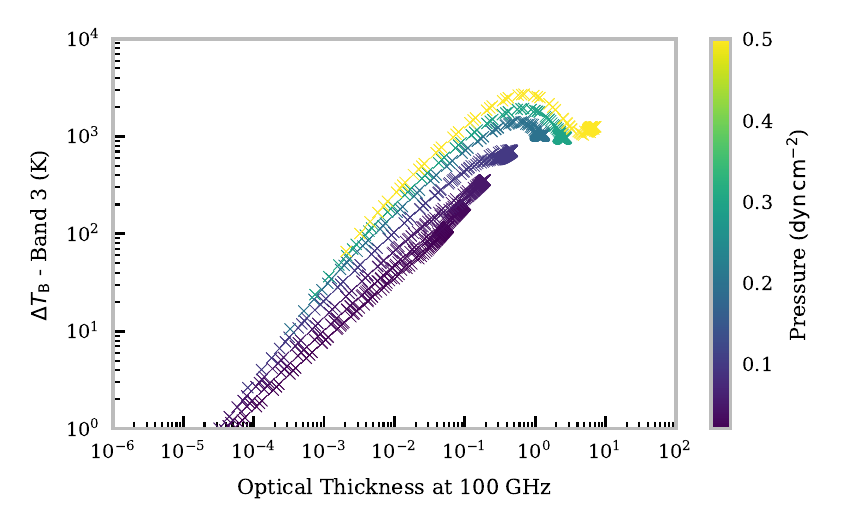}
	\end{subfigure}
	\caption{Figure showing the brightness temperature difference across the ALMA Band~3 sub-band spectra created by the set of isothermal models (left panel), and multi-thermal models (right panel).
	To give an estimate of the spectral gradient the uncertainty of the brightness temperature must be less than the brightness temperature difference across the spectrum.}
	\label{fig:diff}
\end{figure*}

Estimating the optical thickness regime using the logarithmic spectral gradient of the millimetre continuum will require suitably precise measurements of the brightness temperature across the ALMA sub-band. 
	Ideally for the gradient of the logarithmic spectrum to be calculated the uncertainty in the brightness temperature should be significantly less than the brightness temperature difference across the sub-band spectrum.
	Higher precision will thus be necessary when the brightness temperature is very low, or when the spectral gradient tends towards 0 for fully optically thick material. 
	In Figure~\ref{fig:diff} we show the brightness temperature difference across ALMA Band~3 for both the sets of isothermal and multi-thermal models used in this study. 
As may be expected, very low optical thickness, and therefore very low brightness temperature models, will require significantly better precision in the brightness temperature measurements than models with higher brightness temperatures.

\section{Conclusions}\label{sec:conclusions}

In this study we have discussed how the spectral gradient of the millimetre continuum spectrum may be used as a diagnostic of the optical thickness regime at the centre of the observing band when a thermal bremsstrahlung emission mechanism may be assumed. 
We show the derived expressions for both a logarithmic and linear scale spectral gradient, where we find that through the simplicity of the relationship between spectral gradient and the optical thickness that the logarithmic scale provides a better, simpler diagnostic. 
Through testing the theoretical expression with both isothermal and multi-thermal numerical prominence simulations we find that the spectral gradient can be used to estimate the optical thickness regime at band centre provided that a suitable correction is made to account for a non-constant gaunt factor over the frequency band. 

The results presented here for prominence models are however more generally applicable to any off-limb solar structure.
Enhancement from on-disc structures will follow a similar relationship with the addition of a term dependent on the gradient of the background continuum spectrum.
For on-disc structures, the method will thus require knowledge of both the structure's brightness temperature spectrum, but also of the background brightness temperature spectrum illuminating it from the solar disc.
This may be problematic unless the structure is clearly observed to be above the formation region of the millimetre regime, or where the observed structure is transient in nature.

We find that for an isothermal plasma, if the optical thickness of the emitting material lies within the range $\tau \approx 10^{-1}-10^{1}$, our method may be used to estimate the optical thickness of the material, and that this relationship should always hold. 
However, for a more realistic multi-thermal plasma the relationship will not be able to tell directly the optical thickness for $\tau >1$ where the optical thickness is sufficiently high for the spectral gradient to also be defined by the temperature distribution as opposed to the optical thickness solely.

\begin{acknowledgements}
	AR acknowledges support from a STFC studentship ST/N504075/1.
	NL acknowledges support from STFC grant ST/P000533/1.
	The authors would like to acknowledge the members of ISSI team number 374 ``Solving the Prominence Paradox'' led by NL for helpful discussion around the results and applications presented in this study.
	The authors would also like to thank the anonymous referee for their constructive questions and feedback. 
\end{acknowledgements}

\bibliographystyle{aa}
\bibliography{letter}

\end{document}